\documentclass{amsart}

\usepackage[utf8]{inputenc}
\usepackage{amsmath,amsfonts,amssymb}
\usepackage{mathtools}
\usepackage{color}
\usepackage{paralist}
\usepackage{tikz}
\usepackage[colorinlistoftodos,bordercolor=orange,backgroundcolor=orange!20,linecolor=orange,textsize=scriptsize]{todonotes}
\usepackage{hyperref}
\usepackage{fullpage}

\usepackage{inconsolata}

\newtheorem{theorem}{Theorem}

\newtheorem{corollary}[theorem]{Corollary}

\newcommand{\etc}{etc}
\newcommand{\ie}{i.e.}
\newcommand{\eg}{e.g.}
\newcommand{\cf}{cf.}
\newcommand{\eps}{\varepsilon}
\newcommand{\R}{\mathbb{R}}
\newcommand{\Pcal}{\mathcal{P}}
\newcommand{\Qcal}{\mathcal{Q}}

\newcommand{\Coq}{\textsc{Coq}}

\newcommand{\leqlex}{\mathrel{\leq_\mathrm{lex}}}

\newcommand{\CoqPolyhedra}{\textrm{Coq-Polyhedra}}
\newcommand{\MathComp}{Mathematical Components}
\newcommand{\MathCompShort}{MathComp}
\newcommand{\scalar}[2]{\langle #1, #2 \rangle} 
\newcommand{\LP}{\mathrm{LP}}
\newcommand{\DualLP}{\mathrm{DualLP}}
\newcommand{\trans}[1]{#1^T}
\newcommand{\transinv}[1]{#1^{-T}}
\newcommand{\Pos}{K}
\newcommand{\Neg}{L}
\newcommand{\Ext}{\text{ext}}

\definecolor{dkblue}{rgb}{0,0.1,0.5}
\definecolor{lightblue}{rgb}{0,0.5,0.5}
\definecolor{dkgreen}{rgb}{0,0.4,0}
\definecolor{dk2green}{rgb}{0.4,0,0}
\definecolor{dkviolet}{HTML}{932191}
\definecolor{dkpink}{rgb}{1.2,0,1.6}
\definecolor{iden}{HTML}{0332FF}
\definecolor{comment}{HTML}{B12122}

\usepackage{listings}

\lstdefinelanguage{SSR}{
mathescape=true,
texcl=false,
morekeywords=[1]{From, Section, Module, End, Require, Import, Export, Defensive, Function, Variable, Variables, Parameter, Parameters, Axiom, Hypothesis, Hypotheses, Notation, Local, Tactic, Reserved, Scope, Open, Close, Bind, Delimit, Definition, Let, Ltac, Fixpoint, CoFixpoint, Add, Morphism, Relation, Implicit, Arguments, Set, Unset, Contextual, Strict, Prenex, Implicits, Inductive, CoInductive, Record, Structure, Canonical, Coercion, Theorem, Lemma, Corollary, Proposition, Fact, Remark, Example, Proof, Goal, Save, Qed, Defined, Hint, Resolve, Rewrite, View, Search, Show, Print, Printing, All, Graph, Projections, inside, outside, Locate, Maximal},
morekeywords=[2]{forall, exists, exists2, fun, fix, cofix, struct, match, with, end, as, in, return, let, if, is, then, else, for, of, nosimpl},
morekeywords=[3]{Type, Prop},
morekeywords=[4]{pose, set, move, case, elim, apply, clear, hnf, intro, intros, generalize, rename, pattern, after, destruct, induction, using, refine, inversion, injection, rewrite, congr, unlock, compute, ring, field, replace, fold, unfold, change, cutrewrite, simpl, have, gen, generally, suff, wlog, suffices, without, loss, nat_norm, assert, cut, trivial, revert, bool_congr, nat_congr, abstract, symmetry, transitivity, auto, split, left, right, autorewrite},
morekeywords=[5]{by, done, exact, reflexivity, tauto, romega, omega, assumption, solve, contradiction, discriminate},
morekeywords=[6]{do, last, first, try, idtac, repeat},
literate={isn't }{{{\ttfamily\color{dkgreen} isn't }}}1,
showstringspaces=false,
morestring=[b]",
morestring=[d]',
tabsize=2,
extendedchars=true,
sensitive=true,
breaklines=true,
basicstyle=\ttfamily,
captionpos=b,
columns=[l]fullflexible,
identifierstyle={\ttfamily\color{black}},
keywordstyle=[1]{\ttfamily\color{dkviolet}},
keywordstyle=[2]{\ttfamily\color{dkgreen}},
keywordstyle=[3]{\ttfamily\color{lightblue}},
keywordstyle=[4]{\ttfamily\color{dkblue}},
keywordstyle=[5]{\ttfamily\color{red}},
keywordstyle=[6]{\ttfamily\color{dkpink}},
stringstyle=\ttfamily,
commentstyle=\rmfamily,
}

\let\C=\lstinline

\usepackage{etoolbox}
\expandafter\patchcmd\csname \string\lstinline\endcsname{%
  \leavevmode
  \bgroup
}{%
  \leavevmode
  \ifmmode\hbox\fi
  \bgroup
}{}{%
  \typeout{Patching of \string\lstinline\space failed!}%
}

\usetikzlibrary{calc}
\usetikzlibrary{arrows}
\usetikzlibrary{positioning}

\title{A Formalization of Convex Polyhedra based on the Simplex Method}

\thanks{The  authors were partially supported by the programme ``Ingénierie Numérique \& Sécurité'' of ANR, project ``MALTHY'', number ANR-13-INSE-0003, by a public grant as part of the Investissement d’avenir project, reference ANR-11-LABX-0056-LMH, LabEx LMH and by the PGMO program of EDF and FMJH}

\author{Xavier Allamigeon}
\address{INRIA and CMAP, Ecole polytechnique, CNRS, 91128 Palaiseau Cedex, France} 
\email{xavier.allamigeon@inria.fr} 

\author{Ricardo D.~Katz}
\address{CIFASIS-CONICET, Bv. 27 de Febrero 210 bis, 2000 Rosario, Argentina}
\email{katz@cifasis-conicet.gov.ar}


\begin{document}

\begin{abstract}
We present a formalization of convex polyhedra in the proof assistant~\Coq{}. The cornerstone of our work is a complete implementation of the simplex method, together with the proof of its correctness and termination. This allows us to define the basic predicates over polyhedra in an effective way (\ie~as programs), and relate them with the corresponding usual logical counterparts. To this end, we make an extensive use of the Boolean reflection methodology. The benefit of this approach is that we can easily derive the proof of several fundamental results on polyhedra, such as Farkas' Lemma, the duality theorem of linear programming, and Minkowski's Theorem.
\end{abstract}

\maketitle

\section{Introduction}

\paragraph{Context} Convex polyhedra play a major role in many different areas of mathematics and computer science, including algebraic geometry, combinatorics, optimization and operations research, control theory, software verification, compilation and program optimization, constraint solving, \etc. Their success mainly comes from the fact that they provide a convenient tradeoff between expressivity (conjunction of linear inequalities) and tractability. As an illustration of the latter aspect, linear programming, \ie, the optimization of a linear function over linear inequality constraints, can be solved in polynomial time~\cite{Khachiyan80}. 

Among the aforementioned applications of polyhedra, there are some which are critical. For instance, in software verification (\eg, by abstract interpretation), or in control theory, polyhedra are used to compute over-approximations of the set of reachable states or trajectories, in order to provide guarantees on the safety of programs~\cite{CousotHalbwachs78}, and the stability of dynamical systems~\cite{Guglielmi2017}. On the mathematical side, polyhedra are still a very active research subject. For instance, whether linear programming can be solved in strongly polynomial complexity, \ie, by a polynomial time algorithm only doing a number of arithmetic operations bounded by a polynomial in the dimension and the number of constraints, is still an open question, appearing in the list of problems for the $21^\text{th}$ century of Smale~\cite{Smale98}. Another open problem is the existence of a polynomial upper bound on the diameter of polytopes, following the counterexample of \cite{Santos2012} to the linear upper bound conjectured by Hirsch in 1957. In this context, (informal) mathematical software play an increasing role in testing or disproving conjectures (see \eg{}~\cite{Bremner2013}). All this advocates for the formalization of the theory of convex polyhedra in a proof assistant, in order to increase the level of trust in their applications. 

\paragraph{Contributions} In this paper, we present the first steps of a formalization of the theory of convex polyhedra in the proof assistant~\Coq{}. A motivation for using~\Coq{} comes from the longer term objective of formally proving some mathematical results relying on large-scale computation (\eg{}, the counterexample to the Hirsch conjecture given in~\cite{Matschke2015}, a polytope in dimension $20$ with $36\, 425$ vertices). The originality of our approach lies in the fact that our formalization is carried out in an effective way, in the sense that the basic predicates over polyhedra (emptiness, boundedness, membership, etc) are defined by means of \Coq{} programs. All these predicates are then proven to correspond to the usual logical statements. The latter take the form of the existence of certificates: for instance, the emptiness of a polyhedron is shown to be equivalent to the existence of a certificate of inconsistency of the defining system of inequalities (this is known as Farkas' Lemma, see Corollary~\ref{cor:farkas} below for the precise statement). This equivalence between Boolean predicates and formulas living in the kind \C$Prop$ is implemented by using the Boolean reflection methodology, and the supporting tools provided by the \MathComp{} library and its tactic language~\cite{MathComp}. The benefit of the effective nature of our approach is demonstrated by the fact that we easily arrive at the proof of fundamental results on polyhedra, such as several versions of Farkas' Lemma, the duality theorem of linear programming, separation from convex hulls, Minkowski's Theorem, etc.

Our effective approach is made possible by implementing the simplex method inside~\Coq{}, and proving its correctness and termination. Recall that the simplex method is the first algorithm introduced to solve linear programming~\cite{Dantzig51}. Two difficulties need to be overcome to formalize it. On the one hand, we need to deal with its termination. More precisely, the simplex method iterates over the so-called bases. Its termination depends on the specification of a pivoting rule, whose aim is to determine, at each iteration, the next basis. In this work, we have focused on proving that the lexicographic rule~\cite{Dantzig1955} ensures termination. On the other hand, the simplex method is actually composed of two parts. The part previously described, called Phase II, requires an initial basis to start with. Finding such a basis is the purpose of Phase I. It consists in building an extended problem (having a trivial initial basis), and applying to it the Phase~II algorithm. Both phases need to be formalized to obtain a fully functional algorithm. 

We point out that our goal here is \emph{not} to obtain a practically efficient implementation of the simplex method (\eg{}, via the code extraction facility of \Coq{}). Rather, we use the simplex method as a tool in our proofs and, in fact, it turns out to be the cornerstone of our approach, given the intuitionistic nature of the logic in \Coq{}. Thus, we adopt the opposite approach of most textbooks on linear programming where, firstly, theoretical results (like the ones mentioned above) are proven, and then the correctness of the simplex method is derived from them.

The formalization presented in this paper can be found in a library developed by the authors called~\CoqPolyhedra{}. This library is available through a git repository at~{\small \url{https://github.com/nhojem/Coq-Polyhedra}}. The branch {\small\texttt{JAR}} provides the frozen version of the library corresponding to the results presented in this paper.\footnote{It can be downloaded by executing \href{https://github.com/nhojem/Coq-Polyhedra/tree/JAR}{\C$git clone -b JAR https://github.com/nhojem/Coq-Polyhedra.git$}. Instructions for building can be found in the {\small\texttt{README.md}} file.} As mentioned above, our formalization is based on the \MathComp{} library~\cite{MathComp} (\MathCompShort{} for short), which is available at~{\small \url{https://github.com/math-comp/math-comp}}. On top of providing a convenient way to use Boolean reflection, this library contains most of the mathematical tools needed to formalize the simplex method (linear algebra, advanced manipulations of matrices, \etc{}).

\paragraph{Related Work} Our approach has been strongly influenced by the formalization of abstract linear algebra in the \MathComp{} library, which is done in an effective way by exploiting a variant of Gaussian elimination~\cite{Gonthier2011}. 

As far as we know, this is the first formalization of the simplex method in the Calculus of Constructions. In this paradigm, the only work concerning convex polyhedra we are aware of is the implementation of Fourier--Motzkin elimination on linear inequalities in \Coq{}, leading to a proof of Farkas' Lemma~\cite{Sakaguchi2016}. Our work follows a different approach, relying on the theory of linear programming, which has the advantage of providing certificates for the basic predicates over polyhedra. Concerning other families of logics, HOL Light provides a very complete formalization of convex polyhedra~\cite{Harrison2013}, including several important results (Farkas' Lemma, Minkowski's Theorem, Euler--Poincaré formula, \etc). The classical nature of the logic implemented in HOL Light makes it difficult to compare this work with ours. In Isabelle, an implementation of a simplex-based satisfiability procedure for linear arithmetic has been carried out~\cite{Spasic2012}. This is motivated by obtaining a practical and executable code for SMT solving purposes. Here, we are driven by using the simplex method for mathematical proving, which explains why we obtain a completely different kind of formalization.

Our approach has to be distinguished from the previous works aiming at integrating external oracles based on the simplex method~\cite{ObuaTPHOL2005,Besson2007} (and more broadly, on constraint solvers~\cite{BoehmeITP2010,ArmandCPP2011}) into proof assistants. In such approaches, the simplex method is implemented in an ``informal backend'' whose aim is to provide certificates of the validity of some linear inequalities, which are then checked inside the proof assistant. The basic goal is to automate the deduction of some linear inequalities in proof assistants (\eg, the tactics Micromega of \Coq{}). Another possible application is to certify \emph{a posteriori} the computation made by (informal) static analysis tools~\cite{Fouilhe2014}. In contrast, the effectiveness of our approach is based on the implementation of the simplex method \emph{inside} the proof assistant. In this way, the basic properties of polyhedra which are addressed in our work are defined by means of the value returned by the simplex method, and the proofs that these definitions are correct follow from the correctness proof of the simplex method.

\paragraph{Organization of the Paper} In Sect.~\ref{sec:preliminaries}, we introduce basic concepts and results on polyhedra and linear programming. In Sect.~\ref{sec:bases}, we describe the main components of the simplex method, and start its formalization. The lexicographic rule is dealt with in Sect.~\ref{sec:lex}. The two phases of the simplex method are formalized in Sect.~\ref{sec:PhaseII} and~\ref{sec:Phase1}, along with some of the main mathematical results that can be derived from them. Finally, we discuss the outcome of our work in Sect.~\ref{sec:outcome}.

By convention, all \Coq{} definitions, functions, theorems, \etc{} introduced in this work are highlighted in blue when they appear for the first time. This is to distinguish them from the existing material, in particular those brought from the \MathCompShort{} library. We inform the reader that the vast majority of the results described in this paper (especially the ones of Sect.~\ref{sec:bases} to~\ref{sec:Phase1}) are gathered in the file \C$simplex.v$ of \CoqPolyhedra.

\section{Polyhedra, Linear Programming and Duality}\label{sec:preliminaries}

A {\em (convex) polyhedron} is a set of the form 
\[
\Pcal(A,b) \coloneqq \{ x \in \R^n \mid A x \geq b \} \, ,
\]
where $A \in \R^{m \times n}$, $b \in \R^m$, and the inequality $\geq$ is taken entrywise, meaning that $y \geq z$ when $y_i \geq z_i$ for all $i$. In geometric terms, a polyhedron corresponds to the intersection of finitely many halfspaces. A \emph{(affine) halfspace} refers to a set of the form $\{ x \in \R^n \mid \scalar{a}{x} \geq \beta \}$, where $a \in \R^n$, $\beta \in \R$, and $\scalar{\cdot}{\cdot}$ stands for the Euclidean scalar product, \ie{}, $\scalar{x}{y} \coloneqq \sum_i x_i y_i$.

More generally, convex polyhedra can be defined over any ordered field (see \eg{}~\cite{Joswig2016}). This is why our formalization relies on a variable \C$R$ of the type $\C$realFieldType$$ of \MathCompShort{}, whose purpose is to represent an ordered field in which any inequality is decidable. Assume that \C$m$ and \C$n$ are variables of type \C$nat$. The types \C$'M[R]_(m,n)$ and \C$'cV[R]_m$ provided by \MathCompShort{} respectively represent matrices of size $\C$m$ \times \C$n$$ and column vectors of size \C$m$ with entries of type \C$R$. In this paper, we usually omit \C$R$ in the notation of these types, for the sake of readability. The polyhedron associated with \C$A:'M_(m,n)$ and \C$b:'cV_m$ is then defined by means of a Boolean predicate, using the construction \C$pred$ of \MathCompShort{}:
\begin{lstlisting}
Definition |*polyhedron*| A b := [pred x : 'cV_n | (A *m x) >=m b].
\end{lstlisting}
Here, \C$*m$ stands for the matrix product, and \C$>=m$ for the entrywise ordering of vectors: \C$y <=m z$ if and only if \C$y i 0 <= z i 0$ for all \C$i$, where \C$y i 0$ and \C$z i 0$ are respectively the $i$th entry of the vectors \C$y$ and \C$z$ (see \C$vector_order.v$ in \CoqPolyhedra).\footnote{The double index in \C$y i 0$ and \C$z i 0$ comes from the fact that column vectors are encoded as matrices (with one column).}  In this way, \C$polyhedron A b$ is essentially a function from \C$'cV_n$ to \C$bool$, which evaluates whether a vector \C$x$ satisfies \C$(A *m x) >=m b$. The construction \C$pred$ of \MathCompShort{} allows us to use the notation \C$x \in polyhedron A b$, which means that \C$x$ satisfies \C$(A *m x) >=m b$.

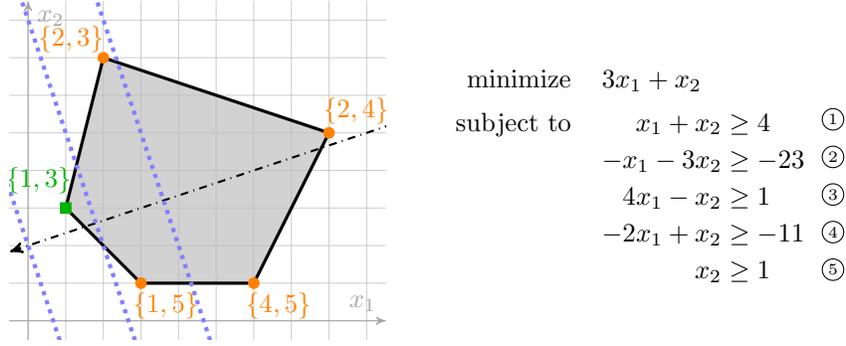
\begin{figure}[t]
\begin{center}
\begin{tikzpicture}[convex/.style={draw=none,fill=lightgray,fill opacity=0.7},convexborder/.style={very thick},point/.style={orange},level_set/.style={blue!50, ultra thick, dotted},>=stealth']
\begin{scope}[scale=.5]
\clip (-0.5,-0.5) rectangle (9.5,8.5);

\draw[help lines,gray!40] (-1,-1) grid (10,10);
\draw[gray!70,->] (-1,0) -- (9.5,0) node[above left] {$x_1$};
\draw[gray!70,->] (0,-1) -- (0,8.5) node[below right] {$x_2$};

\coordinate (v1) at (3,1);
\coordinate (v2) at (6,1);
\coordinate (v3) at (8,5);
\coordinate (v4) at (2,7);
\coordinate (v5) at (1,3);

\fill[convex] (v1) -- (v2) -- (v3) -- (v4) -- (v5) -- cycle;
\draw[convexborder] (v1) -- (v2) -- (v3) -- (v4) -- (v5) -- cycle;

\filldraw[point] (v1) circle (4pt) node[below right=0ex and -1.5ex] {$\{1,5\}$};
\filldraw[point] (v2) circle (4pt) node[below right=0ex and -1.5ex] {$\{4,5\}$};
\filldraw[point] (v3) circle (4pt) node[above right=-0.0ex and -1.3ex] {$\{2,4\}$};
\filldraw[point] (v4) circle (4pt) node[above left=-.3ex and -.8ex] {$\{2,3\}$};
\filldraw[point,green!70!black] (v5) + (-4pt,-4pt) rectangle ++ (4pt,4pt) node[above left=0ex and -.8ex] {$\{1,3\}$};

\coordinate (objective) at ($(3,1)$);
\path let \p1=(objective) in coordinate (level_dir) at ($10*(-\y1, \x1)$);
\draw[dashdotted, thick,<-, >=stealth'] ($(0,2) -0.16*(objective)$) -- ($(0,2)+5*(objective)$) ; 
\draw[level_set] (5,-1) -- + (level_dir);
\draw[level_set] (3,-1) -- + (level_dir);
\draw[level_set] (1,-1) -- + (level_dir);
\end{scope}
\begin{scope}[shift={(8,2)}]
\node[text width=6cm,font=\normalfont] {
\[
\begin{array}{r@{\quad}l}
\text{minimize} & \, 3 x_1 + x_2 \\[1ex]
\text{subject to} & 
\begin{aligned}[t]
x_1 + x_2 & \geq 4 \\ 
-x_1 - 3x_2 & \geq -23 \\ 
4 x_1 - x_2 & \geq 1 \\ 
- 2x_1 + x_2 & \geq -11 \\ 
x_2 & \geq 1
\end{aligned}
\end{array}
\]
};
\begin{scope}[shift={(2.7,0.68)}]
\node[draw,circle,font=\tiny,inner sep=1pt] at (0,0) {$1$};
\node[draw,circle,font=\tiny,inner sep=1pt] at (0,-0.5) {$2$};
\node[draw,circle,font=\tiny,inner sep=1pt] at (0,-1) {$3$};
\node[draw,circle,font=\tiny,inner sep=1pt] at (0,-1.5) {$4$};
\node[draw,circle,font=\tiny,inner sep=1pt] at (0,-2) {$5$};
\end{scope}
\end{scope}
\end{tikzpicture}
\end{center}
\caption{A linear program. The feasible set is depicted in gray. The direction in which the objective function decreases is represented by a dashdotted oriented line, and some level sets (\ie, sets of the form $\{ x\in \R^2 \mid 3 x_1 + x_2 = \alpha \}$, where $\alpha \in \R$) are drawn in blue dotted lines. The optimal basic point is represented by a green square, while the the other basic points are represented by dots. All are annotated with the corresponding bases.}\label{fig:lp}
\end{figure}

\emph{Linear programming} consists in optimizing a linear function $x \in \R^n \mapsto \scalar{c}{x}$ over a polyhedron, such as:
\begin{equation}\tag{$\LP(A,b,c)$} \label{eq:lp}
\begin{array}{r@{\quad}l}
\text{minimize} & \scalar{c}{x} \\[\jot]
\text{subject to} & A x \geq b \, , \; x \in \R^n
\end{array}
\end{equation}
A problem of the form~\ref{eq:lp} is referred to as a \emph{linear program} (see Fig.~\ref{fig:lp} for an example). A vector $x \in \R^n$ satisfying the constraint $A x \geq b$ is a \emph{feasible point} of this linear program. The polyhedron $\Pcal(A,b)$, which consists of the feasible points, is called the \emph{feasible set}. The function $x \mapsto \scalar{c}{x}$ is the \emph{objective} function. The \emph{optimal value} is defined as the infimum of $\scalar{c}{x}$ for $x \in \Pcal(A,b)$. A point $x \in \Pcal(A,b)$ reaching this infimum is called \emph{optimal solution}. When $\Pcal(A,b)$ is not empty, the linear program~\ref{eq:lp} is said to be \emph{feasible}, and its optimal value is either finite, or $-\infty$ (when the quantity $\scalar{c}{x}$ is not bounded from below over $\Pcal(A,b)$). In the latter case, we say that the linear program is \emph{unbounded (from below)}. Finally, when $\Pcal(A,b)$ is empty, the linear program is \emph{infeasible}, and its value is defined to be $+\infty$. 

A fundamental result in linear programming relates the optimal value of the linear program~\ref{eq:lp} with the one of another linear program which is dual to it. In more detail, the \emph{dual linear program} of~\ref{eq:lp} is the following optimization problem:
\begin{equation}\tag{$\DualLP(A,b,c)$} \label{eq:dual_lp}
\begin{array}{r@{\quad}l}
\text{maximize} & \scalar{b}{u} \\[\jot]
\text{subject to} & \trans{A} u = c \, , \; u \geq 0 \, , \; u \in \R^m
\end{array}
\end{equation}
where $\trans{A}$ stands for the transpose of $A$. Notice that \ref{eq:dual_lp} is a linear program as well. Indeed, its constraints can be rewritten into the block system 
\begin{equation}\label{eq:dual}
\begin{pmatrix} \trans{A} \\ -\trans{A} \\ I_m \end{pmatrix} u \geq \begin{pmatrix} c \\ -c \\ 0 \end{pmatrix} \, ,
\end{equation}
where $I_m$ stands for the $m \times m$ identity matrix. Similarly, the maximization problem can be turned into a minimization problem with objective function $x \mapsto \scalar{-b}{x}$. We denote by $\Qcal(A,c)$ the feasible set of~\ref{eq:dual_lp}, and we refer to it as the \emph{dual polyhedron}. Assuming \C$c$ is a variable of type $\C$'cV_n$$ (\ie, representing a vector in $\R^n$), we adopt a specific formalization for this polyhedron, as follows:
\begin{lstlisting}
Definition |*dual_polyhedron*| A c := [pred u : 'cV_m | A^T *m u == c & (u >=m 0)].
\end{lstlisting}
where \C$0$ denotes the zero vector of \C$'cV_m$. Here and below, the notation \C$==$ stands for the Boolean equality, which applies to any type with a decidable equality. As opposed to the dual linear program, \ref{eq:lp} is referred to as the \emph{primal linear program}. The interplay between the primal and dual linear programs is described by the following result:
\begin{theorem}[Strong duality]\label{th:strong_duality}
If one of the programs \ref{eq:lp} or \ref{eq:dual_lp} is feasible, then they have the same optimal value.

In addition, when both are feasible, the optimal value is attained by a primal feasible point $x^* \in \Pcal(A,b)$ and by a dual feasible point $u^* \in \Qcal(A,c)$.
\end{theorem}

In particular, when \ref{eq:dual_lp} is feasible, its optimal value is $+\infty$ if and only if the primal linear program \ref{eq:lp} is infeasible. This holds for any choice of the vector $c$, including $c = 0$. Since $\DualLP(A,b,0)$ obviously admits $u = 0$ as a feasible point, we readily obtain a characterization of the emptiness of the polyhedron $\Pcal(A,b)$, which is known as Farkas' Lemma:
\begin{corollary}[Farkas' Lemma]\label{cor:farkas}
The polyhedron $\Pcal(A,b)$ is empty if, and only if, the optimal value of $\DualLP(A,b,0)$ is $+\infty$, or, equivalently, there exists $d \in \R^m$ such that $d \geq 0$, $\trans{A} d = 0$ and $\scalar{b}{d} > 0$.\footnote{When such a $d$ exists, the value of the objective function of $\DualLP(A,b,0)$ can be made arbitrarily large by considering points of the form $\alpha d$, for $\alpha \geq 0$, which obviously belong to $\Qcal(A,0)$ (due to the latter property, $d$ is called a {\em dual feasible direction}, see Sect.~\ref{subsec:FeasAndFarkas}).}
\end{corollary}

%
%

The first part of Corollary~\ref{cor:farkas} shows a way to formalize the emptiness property of polyhedra in an effective way, \eg, as a program computing the value of $\DualLP(A,b,0)$ inside the proof assistant and comparing it to $+\infty$. This is precisely the approach that we have adopted in this work. As we shall see in Sect.~\ref{sec:outcome}, it also applies to several other properties of polyhedra.


\section{The Three Ingredients of the Simplex Method}\label{sec:bases}

\subsection{Bases and Basic Points}\label{subsec:bases}

To solve the linear program~\ref{eq:lp}, the simplex method iterates over the feasible bases, up to reaching one corresponding to an optimal solution or concluding that the optimal value is $-\infty$. A \emph{basis} is a subset $I$ of $\{1,\ldots ,m\}$ with cardinality $n$ such that the square matrix $A_I$, formed by the rows $A_i$ of $A$ indexed by $i \in I$, is invertible. Each basis $I$ is associated with a \emph{basic point} $x^I$ defined as:
\[
x^I \coloneqq (A_I)^{-1} b_I \, .
\]
The basis $I$ is said to be \emph{feasible} when the point $x^I$ is feasible. It is said to be \emph{optimal} when $x^I$ is an optimal solution of the linear program. We refer to Fig.~\ref{fig:lp} for an illustration.

In geometric terms, a basis corresponds to a set of $n$ hyperplanes $A_i x = b_i$ which intersect in a single point. The basis is feasible when this point belongs to the feasible set $\Pcal(A,b)$. It can be shown that the feasible basic points precisely correspond to the vertices, \ie, the $0$-dimensional faces, of the polyhedron $\Pcal(A,b)$ (albeit we do not prove nor need this property in this paper). 

The formalization of bases and feasible bases is performed by introducing three layers of types. We start with a type corresponding to \emph{prebases}, \ie, subsets of $\{1,\ldots ,m\}$ with cardinality $n$. This is implemented by using a dependant pair formed by a subset \C$I : {set 'I_m}$ and a proof of the equality \C$#|I| == n$:
\begin{lstlisting}
Inductive |*prebasis*| := !*Prebasis*! (I : {set 'I_m}) of (#|I| == n).
\end{lstlisting}
Here, \C$'I_m$ stands for the finite subtype of naturals \C$i:nat$ such that \C$i < m$ (\cf~Interface \C$finType$ of \MathCompShort{}). 
A term \C$I$ of type \C${set 'I_m}$ represents a finite set of elements of type \C$'I_m$, and \C$#|I|$ corresponds to its cardinality. 

Defining bases then requires us to deal with submatrices of the form $A_I$. This is the purpose of the library \C$row_submx.v$ of~\CoqPolyhedra{}, where we define:
\begin{lstlisting}
Definition |*row_submx*| (p : nat) (Q : 'M_(m,p)) (I : {set 'I_m}) := 
	(\matrix_(i < #|I|, j < p) Q (enum_val i) j) : 'M_(#|I|, p).
\end{lstlisting}
In this definition, \C$(\matrix_(i < q,j < p) Expr(i,j)) : 'M_(q,p)$ is the matrix whose \C$(i,j)$ entry is \C$Expr(i,j)$, and the function \C$enum_val$ applied to \C$i:'I_(#|I|)$ retrieves the \C$i$th element of \C$I$. When \C$I$ has cardinality \C$n$ and \C$Q$ has type \C$'M_(m,n)$, the submatrix \C$row_submx Q I$ does not have type \C$'M_n$, \ie, that of square matrices of size $\C$n$ \times \C$n$$. Indeed, in \MathCompShort{}, matrices are defined using dependent types (depending on the size). Therefore, the types \C$'M_n$ and \C$'M_(#|I|,n)$ are distinct, so we use the \MathCompShort{} function \C$castmx$ to explicitly do the glueing job:
\begin{lstlisting}
Definition |*matrix_of_prebasis*| (p : nat) (Q : 'M_(m,p)) (I : prebasis) := 
	castmx (prebasis_card I, erefl p) (row_submx Q I) : 'M_(n,p).
\end{lstlisting}
where \C$|*prebasis_card*| I$ is a proof of the fact that \C$#|I| = n$ and \C$erefl p$ of the fact that \C$p = p$. Assuming the variables \C$A:'M_(m,n)$ and \C$b:'cV_m$ have been previously declared, the type representing bases is then defined by:
\begin{lstlisting}
Inductive |*basis*| := !*Basis*! (I : prebasis) of (matrix_of_prebasis A I) \in unitmx.
\end{lstlisting}
where the predicate \C$unitmx$ represents the set of matrices of \C$'M_n$ which are invertible. The basic point associated with a basis is determined by the function: 
\begin{lstlisting}
Definition |*point_of_basis*| (I : basis) :=  (invmx (matrix_of_prebasis A I)) *m (matrix_of_prebasis b I).
\end{lstlisting}
where \C$invmx Q$ returns the inverse of the matrix \C$Q$. From this, we can define the type of feasible bases:
\begin{lstlisting}
Inductive |*feasible_basis*| := !*FeasibleBasis*! (I : basis) of point_of_basis I \in polyhedron A b.
\end{lstlisting}

\subsection{Reduced Costs}\label{subsec:reduced_costs}

As mentioned above, the simplex method iterates over the feasible bases until reaching one corresponding to an optimal solution or concluding that the optimal value is $-\infty$. This stopping criterion is determined using the so-called \emph{reduced cost vector}. The reduced cost vector associated with the basis $I$ is defined as $u^I \coloneqq\transinv{A_I} c$, where $\transinv{A_I}$ denotes the inverse of the transpose matrix of $A_I$. On the \Coq{} side, assuming \C$c$ is a variable of type \C$'cV_n$, this leads to:
\begin{lstlisting}
Definition |*reduced_cost_of_basis*| (I : basis) := (invmx (matrix_of_prebasis A I)^T) *m c : 'cV_n.
\end{lstlisting}
where \C$Q^T$ stands for the transpose of the matrix \C$Q$. When $u^I \geq 0$ and $I$ is feasible, the associated basic point is optimal:
\begin{lstlisting}
Lemma |*optimal_cert_on_basis*| (I : feasible_basis) : (reduced_cost_of_basis I) >=m 0 ->
	forall y, y \in polyhedron A b -> '[c, point_of_basis I] <= '[c,y].
\end{lstlisting}
Here, the notation \C$'[.,.]$ corresponds to the scalar product $\scalar{\cdot}{\cdot}$:
\begin{lstlisting}
Definition |*vdot*| u v := \sum_i (u i 0) * (v i 0).?\vskip.5ex?
Notation "''[' u , v ]" := vdot u v.
\end{lstlisting}
We refer to the file \C$inner_product.v$ in \CoqPolyhedra{} for properties of \C$vdot$.

The proof of \C$Lemma optimal_cert_on_basis$ relies on two properties. The first one, known as \emph{weak duality}, states that the value of the objective function of~\ref{eq:lp} at any feasible point is greater than or equal to the value of the objective function of~\ref{eq:dual_lp} at any dual feasible point:
\begin{lstlisting}
Lemma |*weak_duality*| x u :	x \in polyhedron A b -> u \in dual_polyhedron -> '[c, x] >= '[b, u].
\end{lstlisting}
The second property is the following geometric interpretation of the reduced cost vector in terms of the dual linear program~\ref{eq:dual_lp}. Given a basis $I$, we introduce the \emph{extended reduced cost vector} $\bar u^I \in \R^m$, which is defined by $\bar u^I_i \coloneqq u^I_i$ if $i \in I$, and $\bar u^I_i \coloneqq 0$ otherwise. On the \Coq{} side, this extended vector is built by the function \C$|*ext_reduced_cost_of_basis*|: basis -> 'cV_m$ (whose exact definition is omitted here for reasons of space). Then, when $u^I \geq 0$, we can show that $\bar u^I$ is a dual feasible point:
\begin{lstlisting}
Lemma |*ext_reduced_cost_dual_feasible*| (I : basis) : 
  let: u := reduced_cost_of_basis I in
  	(u >=m 0) = (ext_reduced_cost_of_basis I \in dual_polyhedron A c).
\end{lstlisting}
Moreover, $\bar u^I$ has the same objective value as the basic point $x^I$:
\begin{lstlisting}
Lemma |*eq_primal_dual_value*| (I : basis) :	'[c, point_of_basis I] = '[b, ext_reduced_cost_of_basis I].
\end{lstlisting}
Thus, as a consequence of \C$Lemma weak_duality$, we conclude that the basic point $x^I$ reaches the optimal value of~\ref{eq:lp} when $u^I \geq 0$.

Another consequence of this interpretation of the reduced cost vector is that the termination of the simplex method immediately provides a proof of the duality theorem of linear programming: the reduced cost vector yields to the dual feasible point $\bar u^I$ which has the same value as the primal feasible point $x^I$.

\subsection{Pivoting}\label{subsec:pivoting}

\emph{Pivoting} refers to the operation of moving from a feasible basis to a ``better'' one, chosen according to what is known as the \emph{pivoting rule}. More precisely, when the reduced cost vector $u^I$ associated with the current feasible basis $I$ does not satisfy the condition $u^I \geq 0$, the pivoting rule selects an index $i \in \{1,\ldots ,n\}$ such that $u^I_i < 0$. The $i$th element of $I$, which in what follows we denote by $l$, is called the \emph{leaving variable}. Then, the direction vector $d \coloneqq (A_I)^{-1} e_i$ (where $e_i$ is the $i$th vector of the canonical base of $\R^n$) is built:
\begin{lstlisting}
Definition |*direction*| (I : basis) (i : 'I_n) := 
	let: ei := (delta_mx i 0):'cV_n in
		(invmx (matrix_of_prebasis A I)) *m ei.
\end{lstlisting}
(here \C$delta_mx i j$ is the matrix with a $1$ in the \C$(i,j)$ entry and $0$ elsewhere) along which the objective function $x \mapsto \scalar{c}{x}$ decreases:
\begin{lstlisting}
Lemma |*direction_improvement*| c (I : basis) (i : 'I_n) :
	(reduced_cost_of_basis I) i 0 < 0 -> '[c, direction I i] < 0.
\end{lstlisting}
As a consequence, the simplex method moves along the halfline $\{x^I + \lambda d \mid \lambda \geq 0\}$ in order to decrease the value of the objective function. When $d$ is a \emph{feasible direction}, meaning that $A d \geq 0$, this halfline is entirely contained in the polyhedron $\Pcal(A,b)$. In this case, we can easily show that the linear program~\ref{eq:lp} is unbounded:
\begin{lstlisting}
Definition |*feasible_dir*| A := [pred d | (A *m d) >=m 0].?\vskip.5ex? 
Lemma |*unbounded_cert_on_basis*| (I : feasible_basis) (i : 'I_n) :
	feasible_dir A (direction I i) -> (reduced_cost_of_basis I) i 0 < 0 ->
		forall M, exists x, (x \in polyhedron A b) /\ ('[c,x] < M).
\end{lstlisting}
In contrast, if $d$ is not a feasible direction, moving along the halfline above makes the simplex method eventually hit the boundary of one of the halfspaces $\{ x \in \R^n \mid A_j x \geq b_j \}$ delimiting $\Pcal(A,b)$. This happens precisely when $\lambda$ reaches the threshold value $\bar \lambda$ defined by:
\begin{equation}
\bar \lambda \coloneqq \min_j \biggl\{ \frac{b_j - A_j x^I}{A_j d} \mid A_j d < 0 \biggr\} \, . \label{eq:mingap}
\end{equation}
The indices attaining the minimum in Eq.~\eqref{eq:mingap} correspond to the halfspaces which are hit. Then, the pivoting rule selects one of them, say $j$, which is called the \emph{entering variable}, and the next basis is defined as $J \coloneqq (I \setminus \{l\}) \cup \{j\}$. In this way, it can be shown that $J$ is a feasible basis, and that $\scalar{c}{x^J}\leq \scalar{c}{x^I}$.


The major difficulty arising in this scheme is the possibility that $\bar \lambda = 0$, or, equivalently, that several bases correspond to the same basic point. Such bases are said to be \emph{degenerate}, and constitute the only obstacle to the termination of the simplex method. In the presence of degenerate bases, the pivoting rule needs to choose carefully the entering and leaving variables in order to avoid cycling over degenerate bases. Our formalization of the simplex method is based on a rule having this property, called the {\em lexicographic rule}~\cite{Dantzig1955}, which is described in the next section. 

\section{Lexicographic Pivoting Rule}\label{sec:lex}

\begin{figure}
\begin{center}
\begin{tikzpicture}[general/.style={fill=lightgray,draw=black!80,thin,fill opacity=.4},vertex/.style={orange}]

\begin{scope}[shift={(-3,0)}, scale=.75,gray!50!black,,>=stealth']
\draw[->] (0,0,0) -- (1,0,0) node[right] {$x_2$};
\draw[->] (0,0,0) -- (0,0,1) node[below left=-1ex] {$x_1$};
\draw[->] (0,0,0) -- (0,1,0) node[above] {$x_3$};	
\end{scope}

\begin{scope}
\draw[general] (-1,-1,-1)--(1,-1,-1)--(1,-1,1)--(-1,-1,1) -- cycle;
\draw[general] (-1,-1,1)--(-1,-1,-1)--(0,1,0) -- cycle;
\draw[general] (-1,-1,-1)--(0,1,0)--(1,-1,-1) -- cycle;
\draw[general] (1,-1,-1)--(1,-1,1)--(0,1,0) -- cycle;
\draw[general] (1,-1,1)--(-1,-1,1)--(0,1,0) -- cycle;

\filldraw[vertex] (0,1,0) circle (.4ex);
\end{scope}

\begin{scope}[shift={(5,0)}]
\coordinate (v0) at (69/64,-33/32,-73/64);
\coordinate (v1) at (69/64,-33/32,67/64);
\coordinate (v2) at (-1/64,37/32,-3/64);
\coordinate (v3) at (-11/64,37/32,-3/64);
\coordinate (v4) at (-81/64,-33/32,-73/64);
\coordinate (v5) at (-81/64,-33/32,67/64);

\draw[general] (v0) -- (v1) -- (v5) -- (v4) -- cycle;
\draw[general] (v3) -- (v4) -- (v5) -- cycle;
\draw[general] (v0) -- (v2) -- (v3) -- (v4) -- cycle;
\draw[general] (v0) -- (v1) -- (v2) -- cycle;
\draw[general] (v1) -- (v2) -- (v3) -- (v5) -- cycle;

\filldraw[vertex,green!80!black] (v2) circle (.4ex);
\filldraw[vertex,green!80!black] (v3) circle (.4ex);
\end{scope}
\end{tikzpicture}	
\end{center}

\caption{Left: a pyramid with square base, defined by the inequalities $2x_2 - x_3 \geq -1$, $2x_1 - x_3 \geq -1$, $-2x_2-x_3 \geq -1$, $-2x_1 - x_3 \geq -1$, and $x_3 \geq -1$. Its apex (represented by an orange dot) is the basic point associated with four different bases. Each of these bases corresponds to a set composed of three distinct defining halfspaces containing this point in their boundary. Right: a perturbation of this polyhedron, as described in Sect.~\ref{sec:lex}, with $\eps = 0.5$. In this case, no basis is degenerate, since each basic point (in particular the two on the top, depicted by green dots) belongs to the boundary of precisely three defining halfspaces.}\label{fig:perturbation}
\end{figure}
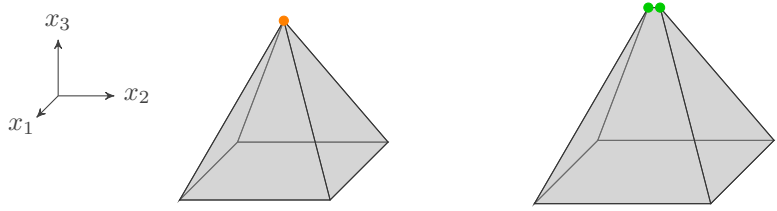

In informal terms, the lexicographic rule acts as if the vector $b$ was replaced by a perturbed vector $\tilde b$ defined by $\tilde b_i \coloneqq b_i - \eps^i$, where $\eps$ is a small positive parameter (here $\eps^i$ is the usual exponentiation). The advantage of perturbing $b$ in such a way is that there is no degenerate basis anymore, see Fig.~\ref{fig:perturbation} for an example. 
In the formalization, which is carried out in \C$Section |*Lexicographic_rule*|$ of \C$simplex.v$, we have chosen to use a symbolic perturbation scheme in order to avoid dealing with numerical values for $\eps$. Indeed, finding how small $\eps$ should be chosen is tedious, and this would make proofs unnecessarily complicated and hard to maintain. We first describe the principle of the symbolic perturbation scheme, and then present our formalization of the lexicographic perturbation rule.

%

\subsection{Symbolic perturbation scheme}\label{subsec:PertScheme}

The symbolic perturbation scheme relies on the fact that we only need to manipulate perturbed real quantities in which the perturbation in $\eps$ is of order less than or equal to $m$. This means that a real number $v$ is perturbed into a quantity of the form:
\[
v + v_1 \eps + \dots + v_m \eps^m \, ,
\]
where $v_1, \dots, v_m \in \R$ and $\eps > 0$. Such a perturbation is symbolically represented by the row vector $(v, v_1 , \dots, v_m) \in \R^{1 \times (1+m)}$.  


We can easily lift the standard operations over such perturbed reals to their symbolic encoding. The addition of two perturbed quantities $\tilde v = \sum_{i = 0}^m v_i \eps^i$ and $\tilde w = \sum_{i = 0}^m w_i \eps^i$ is represented by the entrywise addition $(v_0 + w_0, \dots, v_m + w_m)$ of the corresponding vectors $(v_0, \dots, v_m)$ and $(w_0, \dots, w_m)$. The multiplication of $\tilde v$ by a (nonperturbed) scalar $\lambda \in \R$ corresponds to the multiplication of $(v_0, \dots, v_m)$ by $\lambda$, \ie, $(\lambda v_0, \dots, \lambda v_m)$. The total order~$\leq$ between perturbed reals is given by the lexicographic order $\leqlex$ between the corresponding vectors: $\tilde v \leq \tilde w \iff (v_0, \dots, v_m) \leqlex (w_0, \dots, w_m) $, the latter equivalence being valid as soon as $\eps$ is sufficiently small. 
 By extension, a matrix $x \in \R^{p \times (1+m)}$ can be seen as representing the perturbed vector $\tilde x \in \R^p$, where $\tilde x_i$ is the perturbed real represented by the $i$th row of $x$ (which is of size $1+m$). In this case, the first column of $x$ corresponds to the vector of $\R^p$ being perturbed.

As an example, the perturbed vector $\tilde b$ described above is represented by the row block matrix $\begin{pmatrix} b \; & \; -I_m \end{pmatrix} \in \R^{m \times (1+m)}$ because its $i$th row is the vector $(b_i, 0, \dots, 0, -1, 0, \dots, 0)$, so it represents the quantity $b_i - \eps^i$, as desired. Using the facilities of \MathCompShort{}, this row block matrix is implemented as follows:
\begin{lstlisting}
Definition |*b_pert*| := (row_mx b -(1%:M)):'M_(m,1+m).
\end{lstlisting}
where \C$row_mx U V$ is the row block matrix obtained by concatenating the matrices \C$U$ and \C$V$, and \C$1

As a consequence, the symbolic perturbation scheme leads us to consider polyhedra over $\R^{n \times (1+m)}$ instead of $\R^n$, replacing the entrywise order $\leq$ between elements of $\R^m$ that we previously used to define $\Pcal(A,b)$ by the row-wise order over elements of $\R^{m \times (1+m)}$, where rows are compared according to the lexicographic order $\leqlex$. In more detail, the perturbed polyhedron 
\[
\Pcal(A, \tilde b) = \bigl\{ x \in \R^n \mid \forall i \in [m] \, , \; A_i x \leq \tilde b_i \bigr\} \, ,
\]
is encoded by the set of $x \in \R^{n \times (1+m)}$ satisfying the following system of lexicographic inequalities:
\begin{equation}\label{eq:lex_system}
A_i x \leqlex (b_i, 0, \dots, 0, -1, 0, \dots, 0) \, , \quad \text{for all} \; i \in [m] \, .
\end{equation}
We point out that the matrix $A \in \R^{m \times n}$ is not perturbed, so that $A_i x$ stands for the standard matrix multiplication of the $i$th row of $A$ by $x$, yielding a row vector of size $1+m$.

\subsection{Lex-feasible bases, and non-degeneracy}\label{subsec:non-degeneracy}

In the perturbed setting, the notion of bases defined in Sect.~\ref{subsec:bases} remains unchanged, as it solely depends on the matrix $A$. In contrast, the notion of feasibility, and subsequently, of feasible bases, has now to be understood in the sense of the inequality constraints given in Eq.~\eqref{eq:lex_system}. In other words, in the perturbed setting the basic point associated with the basis $I$ is defined by:
\begin{lstlisting}
Definition |*point_of_basis_pert*| (I : basis) :=
  (invmx (matrix_of_prebasis A I)) *m (matrix_of_prebasis b_pert I) : 'M_(n,1+m).
\end{lstlisting}
Logically, its type corresponds to elements of $\R^{n \times (1+m)}$. The lemma:
\begin{lstlisting}
Lemma |*rel_points_of_basis*| (I : basis) : point_of_basis I = col 0 (point_of_basis_pert I).
\end{lstlisting}
shows that \C$point_of_basis_pert I$ represents a perturbation of the nonperturbed basic point \C$point_of_basis I$, in the sense that the first column of the former coincides with the latter. 

The bases that are feasible in the perturbed setting, which we refer to as \emph{lex-feasible bases} to avoid confusion, are the bases \C$I:basis$ such that the associated symbolically perturbed basic point \C$point_of_basis_pert I$ satisfies the constraints given in Eq.~\eqref{eq:lex_system}: 
\begin{lstlisting}
Definition |*is_lex_feasible*| (I : basis) :=   
  [forall i, ((row i A) *m (point_of_basis_pert I)) >=lex (row i b_pert)].?\vskip.5ex?
Inductive |*lex_feasible_basis*| := !*LexFeasibleBasis*! (I : basis) of is_lex_feasible I.
\end{lstlisting}
where \C$>=lex$ is the lexicographic ordering over row vectors (see \C$vector_order.v$ in \CoqPolyhedra). We observe that any lex-feasible basis is feasible:
\begin{lstlisting}
Lemma |*lex_feasible_basis_is_feasible*| (I : lex_feasible_basis) : is_feasible I.
\end{lstlisting}
as a consequence of \C$Lemma rel_points_of_basis$ and the fact that \C$(u <=lex v) -> u 0 0 <= v 0 0$. We point out that, in general, not every feasible basis is lex-feasible.

As stated previously, the benefit of the symbolic perturbation scheme is to remove degenerate bases, \ie, no two distinct bases correspond to the same perturbed basic point:
\begin{lstlisting}
Lemma |*eq_pert_point_imp_eq_bas*| (I I' : basis) :
	point_of_basis_pert I = point_of_basis_pert I' -> I == I'.
\end{lstlisting}
This turns out to be an immediate consequence of the fact that the \C$(1+j)$th column of \C$point_of_basis_pert I$ is nonzero if, and only if, \C$j$ belongs to \C$I$:
\begin{lstlisting}
Lemma |*col_point_of_basis_pert*| (I : basis) (j : 'I_m) :
	(col (rshift 1 j) (point_of_basis_pert I) != 0) = (j \in I).
\end{lstlisting}
Here and below, given \C$p:nat$ and \C$k:'I_q$, the term \C$rshift p k$ provided by \MathCompShort{} stands for the element of type \C$'I_(p+q)$ corresponding to the integer \C$p+k$. 
The previous result can be proved as follows. Since the matrix $A_I$ is invertible, the \C$(1+j)$th column of \C$point_of_basis_pert I$ is nonzero if, and only if, the \C$(1+j)$th column of \C$matrix_of_prebasis b_pert I$ is. In consequence, \C$Lemma col_point_of_basis_pert$ can be derived from the following lemma:
\begin{lstlisting}
Lemma |*col_b_pert*| (I : prebasis) (j : 'I_m) :
  (col (rshift 1 j) (matrix_of_prebasis b_pert I) != 0) = (j \in I).
\end{lstlisting}
which directly results from the way the vector \C$b_pert$ is defined. Indeed, the \C$(1+j)$th column of the matrix \C$matrix_of_prebasis b_pert I$ has only zero entries, except in the case where \C$j \in I$ (in this case, the entry corresponding to the index of \C$j$ in \C$I$ is $-1$).

\subsection{Pivoting over lex-feasible bases}

The simplex method with the lexicographic rule iterates over lex-feasible bases. By \C$Lemma lex_feasible_basis_is_feasible$, this actually corresponds to iterating over a certain subset of the feasible bases of the original polyhedron $\Pcal(A,b)$. 

The formalization of the lexicographic rule closely follows the description of the pivoting step given in Sect.~\ref{subsec:pivoting}. The only difference is that some of the quantities there are perturbed, and so they are replaced by their symbolic encoding. Note that, in contrast, the definition and properties of the reduced cost and direction vectors of Sect.~\ref{subsec:reduced_costs} and~\ref{subsec:pivoting} are still valid in the perturbed setting, as the matrix $A$ and the vector $c$ are not perturbed. 

We now assume that two variables \C$I:lex_feasible_basis$ and \C$i:'I_n$ have been declared, and we make the following assumptions:\footnote{The command \C$Hypothesis h:P$ is synonymous to the command \C$Variable h:P$ that is used for local declarations when \C$h$ is an identifier and \C$P$ is a proposition. This means that $h$ is a proof term of \C$P$. The hypothesis is then added to list of hypotheses (or, equivalently, of variables) of the subsequent statements.}
\begin{lstlisting}
Hypothesis |*leaving*|: (reduced_cost_of_basis I) i 0 < 0.?\vskip.5ex? 
Hypothesis |*infeasible_dir*|: ~~(feasible_dir A (direction I i)).
\end{lstlisting}
where \C$~~b$ stands for the Boolean negation of \C$b$ of type \C$bool$. These two assumptions ensure that the current basis \C$I$ is not optimal, and that the linear program is not unbounded along the direction \C$direction I i$, see Sect.~\ref{subsec:pivoting}. For the sake of readability, we simply denote \C$direction I i$ by \C$d$ in the rest of the section.

Our aim is to determine an entering variable. In the symbolic perturbation scheme, every ratio appearing in Eq.~\eqref{eq:mingap} turns out to be a row vector encoding a perturbed real:
\begin{lstlisting}
Definition |*lex_gap*| (j : 'I_m) := 
  let: x_pert := point_of_basis_pert I in
    ((A *m d) j 0)^-1 *: ((row j b_pert) - ((row j A) *m x_pert)) : 'rV_(1+m).
\end{lstlisting}
where \C$'rV_(1+m)$ stands for the type of row vectors of size $1+m$, and \C$*:$ for the product of a scalar by a matrix (\ie, if \C$lamb:R$ and \C$Q:'M_(q,p)$, then \C$lamb *: Q$ is the matrix of type \C$'M_(q,p)$ whose \C$(i,j)$ entry is \C$lamb * (Q i j)$). In order to obtain in the perturbed setting the analog of the threshold value $\bar \lambda$ defined in Eq.~\eqref{eq:mingap}, we determine the minimum of these ratios in the lexicographic sense, using the function \C$|*lex_min_seq*| S$ introduced in the file \C$vector_order.v$ of \CoqPolyhedra. The entering variable is then computed as follows:
\begin{lstlisting}
Definition |*lex_ent_var_nat*| :=
  let: J := [seq j <- (enum 'I_m) | (A *m d) j 0 < 0] in
  let: min_gap := lex_min_seq [seq lex_gap j | j <- J] in
    find (fun j => (j \in J) && (min_gap == lex_gap j)) (enum 'I_m).
\end{lstlisting}
where \C$enum 'I_m$ represents a duplicate-free list of all the \C$j \in 'I_m$, and the \MathCompShort{} function \C$find P S$ returns the index of the first item in the sequence \C$S$ for which the Boolean predicate \C$P$ holds, if any. Next, we prove that \C$lex_ent_var_nat$ (of type \C$nat$) is strictly less than \C$m$, which allows us to built the corresponding element of type \C$'I_m$ called \C$|*lex_ent_var*|$. 

We are now ready to build the next basis:
\begin{lstlisting}
Definition |*lex_rule_set*| := lex_ent_var |: (I :\ (enum_val [...] i)).
\end{lstlisting}
Here, the operation \C$(I :\ (enum_val [...] i))$ removes the \C$i$th element of \C$I$ (\ie, the leaving variable $l$, see Sect.~\ref{subsec:pivoting}) from \C$I$, while \C$lex_ent_var |: [...]$ adds \C$lex_ent_var$ to the resulting set (both functions \C$:\$ and \C$|:$ are provided by \MathCompShort{}).

We first prove that \C$lex_rule_set$ has cardinality $n$ and that it constitutes a basis: 
\begin{lstlisting}
Lemma |*lex_rule_card*| : #|lex_rule_set| == n.?\vskip.5ex?
Lemma |*lex_rule_is_basis*| : is_basis (!*Prebasis*! lex_rule_card).?\vskip.5ex?
Definition |*lex_rule_basis*| := !*Basis*! lex_rule_is_basis.
\end{lstlisting}
The proof of \C$Lemma lex_rule_is_basis$ is obtained by noticing that the row of $A$ indexed by \C$lex_ent_var$ is not a linear combination of the rows $A_k$ for $k \in I \setminus \{l\}$, where $l$ is the leaving variable as above. Indeed, by the definition of \C$lex_ent_var$ and \C$lex_ent_var_nat$, the multiplication of this row of $A$ by the direction vector $d$ is a negative real, while $A_k d  = 0$ for all $k \in I \setminus \{l\}$ by the definition of $d$. 

It next remains to show that \C$lex_rule_basis$ is lex-feasible, which, in turn, leads to the definition of the new lex-feasible basis \C$|*lex_rule_lex_bas*|$:
\begin{lstlisting}
Lemma |*lex_rule_lex_feasibility*| : is_lex_feasible lex_rule_basis.?\vskip.5ex?
Definition |*lex_rule_lex_bas*| := !*LexFeasibleBasis*! lex_rule_lex_feasibility.
\end{lstlisting}
We establish \C$Lemma lex_rule_lex_feasibility$ by showing the following identity:
\begin{lstlisting}
Lemma |*lex_rule_rel_succ_points*| : 
	let x := point_of_basis_pert I in
	let x' := point_of_basis_pert lex_rule_basis in
		x' = x + d *m (lex_gap lex_ent_var).
\end{lstlisting}
Recall that the quantity \C$lex_gap lex_ent_var$ is the analog of the threshold $\bar \lambda$ in the perturbed setting. Thus, \C$Lemma lex_rule_rel_succ_points$ means that the new basic point \C$x'$ is obtained from the previous basic point \C$x$ by moving along the direction \C$d$. The lex-feasibility of \C$x'$ essentially follows from the lex-minimality of \C$lex_gap lex_ent_var$ among the quantities \C$lex_gap j$ for \C$j$ ranging over the elements satisfying \C$(A *m d) j 0 < 0$ (we refer to the definition of \C$lex_ent_var_nat$ above). 

Finally, we prove that the analog of the objective function in the perturbed setting is \emph{strictly} decreasing in the lexicographic sense: 
\begin{lstlisting}
Lemma |*lex_rule_dec*| : 
	let: J := lex_rule_lex_bas in
		(c^T *m point_of_basis_pert I) >lex (c^T *m point_of_basis_pert J).
\end{lstlisting}
To prove this lemma, we naturally exploit \C$Lemma lex_rule_rel_succ_points$ and the fact that \C$'[c,d] < 0$ (see \C$Lemma direction_improvement$ in Sect.~\ref{subsec:pivoting}). Above all, we prove: 
\begin{lstlisting}
Lemma |*lex_min_gap_lex_pos*| : lex_gap lex_rule_dec >lex 0.
\end{lstlisting}
While the inequality \C$lex_gap lex_rule_dec >=lex 0$ is easy to obtain, the fact that \C$lex_gap lex_rule_dec != 0$ is established by contradiction, exploiting the absence of degenerate bases in the perturbed setting (\C$Lemma eq_pert_point_imp_eq_bas$).


\section{Phase II of the Simplex Method, and Farkas' Lemma}\label{sec:PhaseII}

\subsection{Phase II} In this section, we present our implementation of \emph{Phase II} of the simplex method. We do it before the one of Phase I because, as we will explain in Sect.~\ref{sec:Phase1}, Phase~II is used in Phase I. Phase II of the simplex method determines the optimal value of~\ref{eq:lp}, supposing that an initial feasible basis \C$bas0:feasible_basis$ is known. \emph{De facto}, this makes the underlying assumption that this linear program is feasible. 

Our formalization of Phase II, which is developed in \C$Section |*Phase2*|$ of \C$simplex.v$, consists in iterating the function \C$lex_rule_lex_bas$ until determining that~\ref{eq:lp} is unbounded (\ie, identifying that the direction vector is feasible), or finding an optimal basis (\ie, identifying that the associated reduced cost vector is nonnegative). Termination is expected to be guaranteed by the fact that no basis can appear twice (as a consequence of \C$Lemma lex_rule_dec$ above) and that the number of bases is finite. 

One might be tempted to start the iteration of \C$lex_rule_lex_bas$ from \C$bas0$. However, albeit feasible, there is no reason for \C$bas0$ to be lex-feasible. Fortunately, it can be shown that, up to reordering the defining inequalities of $\Pcal(A,b)$, we can make \C$bas0$ lex-feasible. More precisely, it can be proved that if $\{m-n+1, \ldots ,m\}$ is a feasible basis, then it is also lex-feasible. Thus, it is enough to reorder the inequalities in such a way that the ones corresponding to indices in \C$bas0$ become the last inequalities of the system. 

Instead of applying permutations on the rows of $A$ and $b$, we choose to apply the inverse permutation on the symbolic perturbation components of \C$b_pert$, and leave the initial problem~\ref{eq:lp} unchanged. As a consequence, we modify the previous definition of \C$b_pert$ as follows:
\begin{lstlisting}
Definition |*b_pert*| := (row_mx b (-(perm_mx s))).
\end{lstlisting}  
where \C$s:'S_m$ represents a permutation of the set $\{1, \ldots ,m\}$, and \C$perm_mx$ builds the corresponding permutation matrix (see the libraries \C$perm$ and \C$matrix$ of \MathCompShort{}). All the previous results remain valid under this change. The only difference is that now they are additionally parametrized by the permutation \C$s$, appearing as a global variable in \C$Section |*Lexicographic_rule*|$. 

Following the discussion above, we define a permutation \C$s0$ (for reasons of space, we omit here its \Coq{} definition) such that the inequalities corresponding to indices in \C$bas0$ become the last ones, and prove that it satisfies the expected result:
\begin{lstlisting}
Lemma |*feasible_to_lex_feasible*| : is_lex_feasible s0 bas0.
\end{lstlisting}

The function performing one iteration of the Phase II algorithm with the lexicographic rule is built as follows:
\begin{lstlisting}
Definition |*basic_step*| (bas : lex_feasible_basis) :=
  let u := reduced_cost_of_basis bas in
    if [pick i | u i 0 < 0] is Some i
    then let d := direction bas i in
      if (@idPn (feasible_dir A d)) is ReflectT infeasible_dir
    	then !*Lex_next_basis*! (lex_rule_lex_bas infeasible_dir)
    	else !*Lex_final*! (!*Lex_res_unbounded*! (bas, i))
    else !*Lex_final*! (!*Lex_res_optimal_basis*! bas).
\end{lstlisting}
Here, the \MathCompShort{} syntax \C$if-is-then-else$ is just syntactic sugar for the \Coq{} pattern matching mechanism. Besides, \C$@idPn (feasible_dir A d)$ returns a proof (namely, \C$infeasible_dir$) of the fact that the direction vector \C$d$ is not feasible, when the Boolean \C$(feasible_dir A d)$ is equal to \C$false$. Finally, when there exists an index \C$i$ such that the Boolean predicate \C$P$ holds, \C$[pick i | P]$ returns \C$Some i$ (the first index with this property is chosen by definition), and \C$None$ otherwise. 

The return type of \C$basic_step$ is an inductive type defined as follows:
\begin{lstlisting}
Inductive |*lex_intermediate_result*| :=
| !*Lex_final*! of lex_final_result
| !*Lex_next_basis*! of (lex_feasible_basis s0).
\end{lstlisting}
This means that \C$basic_step$ returns either a next basis (constructor \C$!*Lex_next_basis*!$), or indicates that the method should stop (constructor \C$!*Lex_final*!$). In the latter case, it carries out a term of type \C$lex_final_result$, which is defined by:
\begin{lstlisting}
Inductive |*lex_final_result*| :=
| !*Lex_res_unbounded*! of (lex_feasible_basis s0) * 'I_n
| !*Lex_res_optimal_basis*! of (lex_feasible_basis s0).
\end{lstlisting}
The constructor \C$!*Lex_res_optimal_basis*!$ is intended to correspond to an optimal lex-feasible basis. The other constructor \C$!*Lex_res_unbounded*!$ provides a lex-feasible basis \C$I$ and an index \C$i:'I_n$ which are expected to identify a feasible direction \C$direction I i$ along which the objective function of~\ref{eq:lp} is unbounded.

The recursive function which iterates the function \C$basic_step$ is the following:
\begin{lstlisting}
Function |*lex_phase2*| bas {measure basis_height bas} :=
  match basic_step bas with
  | !*Lex_final*! final_res => final_res
  | !*Lex_next_basis*! bas' => lex_phase2 bas'
  end.
\end{lstlisting}
It is defined in the framework provided by the library \C$RecDef$ of \Coq{}, see~\cite{Barthe2006}. More precisely, its termination (and subsequently, the fact that \Coq{} accepts the definition) is established by identifying an integer quantity which is strictly decreased every time the function \C$basic_step$ returns a next basis:
\begin{lstlisting}
Definition |*basis_height*| bas :=  #|[set bas':(lex_feasible_basis s0) | 
	(c^T *m (point_of_basis_pert s0 bas')) <lex (c^T *m (point_of_basis_pert s0 bas))]|.
\end{lstlisting}
This quantity represents the number of lex-feasible bases for which the value of the objective function is (lexicographically) strictly less than the value of this function at the current lex-feasible basis. The fact that \C$basis_height$ decreases at each iteration is a direct consequence of \C$Lemma lex_rule_dec$.

Gathering all these components, we finally arrive at the definition of the function implementing Phase~II:
\begin{lstlisting}
Definition |*phase2*| := 
  let: lex_bas0 := !*LexFeasibleBasis*! feasible_to_lex_feasible in
    lex_to_phase2_final_result ((@lex_phase2 s0) c lex_bas0).
\end{lstlisting}
where \C$|*lex_to_phase2_final_result*|$ is a function which transforms an element of type \C$lex_final_result$ into an element of type:
\begin{lstlisting}
Inductive |*phase2_final_result*| :=
| !*Phase2_res_unbounded*! of feasible_basis * 'I_n
| !*Phase2_res_optimal_basis*! of feasible_basis.
\end{lstlisting}
that is, it essentially transforms a lex-feasible basis into a feasible basis. This is possible thanks to \C$Lemma lex_feasible_basis_is_feasible$, see Sect.~\ref{subsec:non-degeneracy}.

We present the correctness specification of \C$phase2$ by means of an adhoc inductive predicate. Such a presentation is idiomatic in the \MathComp{} library. The advantage is that it provides a convenient way to perform case analysis on the result of \C$phase2$. 
\begin{lstlisting}
Inductive |*phase2_spec*| : phase2_final_result -> Type :=
| !*Phase2_unbounded*! (p:feasible_basis * 'I_n) of 
  (reduced_cost_of_basis p.1) p.2 0 < 0 /\ feasible_dir A (direction p.1 p.2) : 
		phase2_spec (!*Phase2_res_unbounded*! p)
| !*Phase2_optimal_basis*! (bas:feasible_basis) of 
  (reduced_cost_of_basis bas) >=m 0 : phase2_spec (!*Phase2_res_optimal_basis*! bas).?\vskip.5ex?
Lemma |*phase2P*| : phase2_spec phase2.
\end{lstlisting} 

To be more precise, \C$Lemma phase2P$ states that when the function \C$phase2$ returns a result of the form \C$!*Phase2_res_unbounded*! (bas,i)$, the pair \C$(bas,i)$ satisfies \C$(reduced_cost_of_basis bas) i 0 < 0$ and \C$feasible_dir A (direction bas i)$. Since these are precisely the hypotheses of \C$Lemma unbounded_cert_on_basis$ of Sect.~\ref{subsec:pivoting}, it follows that \C$(bas,i)$ is a certificate that~\ref{eq:lp} is unbounded. Similarly, if the result of \C$phase2$ is of the form \C$!*Phase2_res_optimal_basis*! bas$, we have \C$(reduced_cost_of_basis bas) >=m 0$, and then by \C$Lemma optimal_cert_on_basis$ of Sect.~\ref{subsec:reduced_costs} we conclude that \C$bas$ is optimal. In other words, \C$Lemma phase2P$ proves that the function \C$phase2$ meets the specification of the terminal case of Phase II that we made at the beginning of this section. 

\subsection{Effective Definition of Feasibility, and Farkas' Lemma.}\label{subsec:FeasAndFarkas}

We can now formalize the notion of feasibility, \ie, the property that a polyhedron $\Pcal(A,b)$ is empty or not, as a Boolean predicate.\footnote{We make a slight abuse of language, since feasibility usually applies to linear programs, or systems of contraints. By extension, we apply it to polyhedra: $\Pcal(A,b)$ is \emph{feasible} if it is nonempty.}
We still assume that the variables \C$A:'M_(m,n)$ and \C$b:'cV_m$ are declared. Following the discussion at the end of Sect.~\ref{sec:preliminaries}, this predicate is defined by means of the function \C$phase2$ applied to $\DualLP(A, b, 0)$. This is a linear program in dimension $m$, with objective function $x \mapsto \scalar{-b}{x}$ (when expressed as a minimization problem) and feasible set given by the predicate \C$dual_polyhedron A 0$. In order to apply \C$phase2$, we need to manipulate this set as a polyhedron of the form $\Pcal(A^\text{dual}, 0)$, defined by the $2n+m$ inequalities in Eq.~\eqref{eq:dual} with $c = 0$:
\begin{lstlisting}
Definition |*dualA*| := col_mx (col_mx A^T (-A^T)) (1%:M) : 'M_((n+n)+m, m).?\vskip.5ex?
Lemma |*dual_coneE*| : dual_polyhedron A 0 =i polyhedron dualA 0.
\end{lstlisting}
where \C$=i$ stands for the extensional equality between Boolean predicates in \MathCompShort{}.

We need an initial dual feasible basis \C$|*dual_feasible_bas0*|$ to run \C$phase2$ on the dual linear program. This basis is built from the indices of the last $m$ inequalities in the system defining $\Pcal(A^\text{dual}, 0)$:
\begin{lstlisting}
Definition |*dual_set0*| := [set (rshift (n+n) i) | i in [set: 'I_m] ].
\end{lstlisting}
where \C$[set: 'I_m]$ represents the set of elements \C$i$ of type \C$'I_m$. 
We easily verify that this forms a basis, since the corresponding submatrix of \C$dualA$ is the identity:
\begin{lstlisting}
Lemma |*dual_set0_card*| : (#|dual_set0| == m)%N.?\vskip.5ex?
Definition |*dual_pb0*| := !*Prebasis*! dual_set0_card.?\vskip.5ex?
Lemma |*dualA_pb0_is_id*| : matrix_of_prebasis dualA dual_pb0 = 1%:M.?\vskip.5ex?
Lemma |*dual_pb0_is_basis*| : is_basis dualA dual_pb0.?\vskip.5ex?
Definition |*dual_bas0*| := Basis dual_pb0_is_basis.
\end{lstlisting}
Moreover, the associated basic point is the identically null point, which ensures that \C$dual_bas0$ is feasible.
\begin{lstlisting}
Lemma [...] : point_of_basis 0 dual_bas0 = 0.?\vskip.5ex?
Lemma |*dual_bas0_is_feasible*| : is_feasible 0 dual_bas0.?\vskip.5ex?
Definition |*dual_feasible_bas0*| := FeasibleBasis dual_bas0_is_feasible.
\end{lstlisting}

Thanks to this, we can now define the property that the polyhedron $\Pcal(A, b)$ is feasible as follows:
\begin{lstlisting}
Definition |*feasible*| := if phase2 dual_feasible_bas0 (-b) is !*Phase2_res_optimal_basis*! _
                       then true else false.
\end{lstlisting}
The correctness of our definition is established by showing that the predicate \C$feasible$ is equivalent to the existence of a point $x$ in $\Pcal(A, b)$. This is presented by means of Boolean reflection, using the \C$reflect$ relation of \MathCompShort{}:\footnote{If \C$P:Prop$ and \C$b:bool$, the statement \C$reflect P b$ means that either \C$b = true$ and $P$, or \C$b = false$ and \C$\~P$ hold. We refer to~\cite{MathComp} for an introduction on the way Boolean reflection is used in \MathCompShort{}.}
\begin{lstlisting}
Lemma |*feasibleP*| : reflect (exists x, x \in polyhedron A b) feasible.
\end{lstlisting}

The feasibility certificate \C$x$ is constructed from the extended reduced cost vector \C$v$ of the optimal basis of $\DualLP(A, b, 0)$ returned by \C$phase2$. As $\Pcal(A^\text{dual}, 0)$ is defined by $2n+m$ inequalities, this extended reduced cost vector is of type \C$'cV_((n+n)+m)$. In the proof of \C$Lemma feasibleP$, the point \C$x$ is then defined as:\begin{lstlisting}
pose x := (dsubmx (usubmx v) - usubmx (usubmx v)).
\end{lstlisting}
Here, \C$usubmx:'cV_(p+q) -> 'cV_p$ and \C$dsubmx:'cV_(p+q) -> 'cV_q$ are \MathCompShort{} functions which respectively return the up and down subvectors of a block vector \C$z:'cV_(p+q)$. Showing that \C$x \in polyhedron A b$ is done by a sequence of rewritings, starting from the fact that \C$v$ belongs to the dual polyhedron associated with $\DualLP(A, b, 0)$, as proved in \C$Lemma ext_reduced_cost_dual_feasible$ of Sect.~\ref{subsec:reduced_costs}.

In a similar way, we prove the following characterization of the emptiness of~$\Pcal(A,b)$, which precisely corresponds to Farkas' Lemma: 
\begin{lstlisting}
Lemma |*infeasibleP*| :	reflect (exists d, dual_feasible_dir A d /\ '[b,d] > 0) (~~feasible). 
\end{lstlisting}
where \C$dual_feasible_dir$ is the analog of the predicate \C$feasible_dir$ in the case of the dual polyhedron:
\begin{lstlisting}
Definition |*dual_feasible_dir*| := [pred d | (A^T *m d == 0) && (d >=m 0)].
\end{lstlisting}
Indeed, \C$~~feasible$ amounts to the fact that \C$phase2$ returns an unboundedness certificate \C$!*Phase2_res_unbounded*! (bas,i)$ for $\DualLP(A, b, 0)$. The emptiness certificate \C$d$ of $\Pcal(A,b)$ is defined as the dual feasible direction \C$direction bas i$. The inequality \C$'[-b,d] < 0$ is directly derived from \C$Lemma direction_improvement$ of Sect.~\ref{subsec:pivoting}.

\section{Complete Implementation of the Simplex Method}\label{sec:Phase1}

\subsection{The Pointed Case} In order to obtain a full formalization of the simplex method, it remains to implement a \emph{Phase~I} algorithm. Its purpose is twofold: 
\begin{inparaenum}[(i)] 
\item determine whether the linear program~\ref{eq:lp} is feasible or not,
\item in the former case, return an initial feasible basis for Phase~II. 
\end{inparaenum}
There is one obstacle to the definition of such a Phase~I algorithm: even if a linear program is feasible, it may not have any feasible basis. For instance, consider the linear program over the variables $x_1, x_2$ which aims at minimizing $x_2$ subject to $-1 \leq x_2 \leq 1$. Its feasible set is a cylinder around the $x_1$-axis, so it does not have any vertex, or, equivalently, basic point. A necessary and sufficient condition for the existence of a feasible basis is that the rank of $A$ be equal to $n$. When this condition is fulfilled, the feasible set $\Pcal(A, b)$ is said to \emph{pointed}. We now describe the Phase~I algorithm under this assumption. This is developed in \C$Section |*Pointed_simplex*|$ of \C$simplex.v$.
\begin{lstlisting}
Hypothesis |*Hpointed*| : (\rank A >= n)%N. 
\end{lstlisting}
(Here, \C$

From the hypothesis on the rank of $A$, we can extract an invertible square submatrix of $A$, which provides an initial basis \C$bas0$ of~\ref{eq:lp} (see the function \C$|*build_row_base*|$ defined in \C$row_submx.v$). Beware that the basis \C$bas0$ is not necessarily a feasible one. As a consequence, we split the inequalities in the system $A x \geq b$ into two complementary groups, $A_\Pos x \geq b_\Pos$ and $A_\Neg x \geq b_\Neg$, where the $\Pos$ is the set of indices $i \in \{1,\ldots ,m\}$ for which the basic point \C$point_of_basis bas0$ does not satisfy the inequality $A_i x \geq b_i$, and $\Neg \coloneqq \{1,\ldots ,m\} \setminus \Pos$. We denote by $p$ the cardinality of the set $K$. 
Phase~I is based on applying Phase~II algorithm to the following ``extended'' problem over the vector $z = (x,y) \in \R^{n+p}$:
\begin{equation}\tag{$\LP_\text{Phase~I}$} \label{eq:phaseI}
\begin{array}{r@{\quad}l}
\text{minimize} & \scalar{e}{y - A_\Pos x} \\[\jot]
\text{subject to} & A_\Pos x \leq b_\Pos + y \, , \; A_\Neg x \geq b_\Neg \, , \; y \geq 0 \, , \; (x,y) \in \R^{n+p} 
\end{array}
\end{equation}
where $e \in \R^p$ stands for the all-$1$-vector. The constraints defining the feasible set of~\ref{eq:phaseI} are gathered into a single system $A_\Ext z \geq b_\Ext$. Similarly, the objective function of~\ref{eq:phaseI} can be written as a linear function $z \mapsto \scalar{c_\Ext}{z}$ of $z = (x,y)$. 
%

The linear program~\ref{eq:phaseI} has two important properties. On the one hand, its optimal value can be bounded (from below) by the quantity $M_\Ext \coloneqq \scalar{e}{-b_\Pos}$: 
\begin{lstlisting}
Definition |*Mext*| := '[const_mx 1, - (row_submx b K)].?\vskip.5ex?
Lemma |*cext_min_value*| z : (z \in polyhedron Aext bext) -> '[cext, z] >= Mext.
\end{lstlisting}
On the other hand, the optimal value of~\ref{eq:phaseI} is equal to $M_\Ext$ if, and only if, the original problem~\ref{eq:lp} is feasible. The ``only if'' implication follows from the following lemma, which also provides a feasibility witness of~\ref{eq:lp}:
\begin{lstlisting}
Lemma |*feasible_cext_eq_min_active*| z :
  z \in polyhedron Aext bext -> '[cext,z] = Mext -> (usubmx z \in polyhedron A b).
\end{lstlisting}
Regarding the ``if'' implication, an infeasibility certificate of~\ref{eq:lp} can be constructed by means of a feasible point $\bar u \in \R^{m+p}$ of the dual of~\ref{eq:phaseI} whose objective value $\scalar{b_\Ext}{\bar u}$ is strictly greater than $M_\Ext$. This certificate is built by the following function:
\begin{lstlisting}
Definition |*dual_from_ext*| (u : 'cV_(m+p)) :=
  \col_i (if i \in K then 1 - (usubmx u) i 0 else (usubmx u) i 0).
\end{lstlisting}
where \C$\col_i Expr(i)$ is the column vector whose $i$th entry is \C$Expr(i)$. As expected, this certificate satisfies:
\begin{lstlisting}
Lemma |*dual_polyhedron_from_ext*| u : 
  u \in dual_polyhedron Aext cext -> dual_feasible_dir A (dual_from_ext u).?\vskip.5ex?
Lemma |*dual_from_ext_obj*| u : '[bext, u] > Mext -> '[b, dual_from_ext u] > 0.
\end{lstlisting}
In this way, by using \C$Lemma infeasibleP$ of Sect.~\ref{subsec:FeasAndFarkas} we readily obtain a proof that~\ref{eq:lp} is infeasible.

Finally, we can build an initial feasible basis \C$|*feasible_bas0_ext*|$ for~\ref{eq:phaseI} by considering the union of \C$bas0$ with the set $\{m+1, \dots, m+p\}$ of the indices of the last $p$ constraints $y \geq 0$ of~\ref{eq:phaseI}. We let the reader check that the associated basic point is $\begin{psmallmatrix} x \\ 0 \end{psmallmatrix} \in \R^{n+p}$, where $x$ is \C$point_of_basis bas0$, and that this point is feasible.

As a consequence, we can apply \C$phase2$ to solve~\ref{eq:phaseI}, starting from the basis \C$feasible_bas0_ext$. 
In this way, we obtain an optimal basis \C$bas$ of~\ref{eq:phaseI}. If the associated basic point $z^*$ satisfies $\scalar{c_\Ext}{z^*} > M_\Ext$, we build an infeasibility certificate of~\ref{eq:lp} using the function \C$dual_from_ext$, as described above. 
Otherwise (\ie, if $\scalar{c_\Ext}{z^*} = M_\Ext$), we construct a feasible basis \C$bas'$ of~\ref{eq:lp} that we use to execute \C$phase2$ on~\ref{eq:lp} and finally obtain its optimal value. This basis \C$bas'$ is provided by the function \C$|*extract_feasible_basis*|$ of \C$simplex.v$. The definition of this function relies on the notion of extreme point. A point $x \in \Pcal(A,b)$ is said to be {\em extreme} if it is not possible to express $x$ as a strict convex combination of two distinct points of $\Pcal(A,b)$. This is formalized in the file \C$polyhedron.v$ of \CoqPolyhedra{} as follows:
\begin{lstlisting}
Definition |*is_extreme*| x (P : pred 'cV_n) := 
  x \in P /\ (forall y z, forall lamb,
             y \in P -> z \in P -> (0 < lamb < 1) -> x = lamb *: y + (1 - lamb) *: z -> x = y /\ x = z).
\end{lstlisting}
(here extremality is defined more generally for a set composed of all the \C$x:'cV_n$ for which a given predicate \C$P$ holds). We prove that the extreme points of $\Pcal(A,b)$ are precisely the feasible basic points of~\ref{eq:lp} (or, intuitively, its vertices). In more detail, we show (see \C$Theorem |*extremality_active_ineq*|$ of \C$polyhedron.v$) that the extremality of $x$ in $\Pcal(A,b)$ is equivalent to the fact that the rank of the submatrix $A_{I(x)}$ is $n$, where $I(x)$ is the set of $i \in [m]$ such that $A_i x = b_i$. As a consequence, every feasible basic point is extreme:
\begin{lstlisting}
Lemma |*feasible_point_of_basis_is_extreme*| (I : feasible_basis) :
  is_extreme (point_of_basis I) (polyhedron A b).
\end{lstlisting}
Conversely, if $x$ is extreme, we can extract a feasible basis $I$ such that $x$ is the associated basic point. The set $I$ is built thanks to the function \C$build_row_base$ (in \C$row_submx.v$) in such a way that $A_I$ is an invertible submatrix of $A_{I(x)}$, and thus of $A$. This extraction procedure is formalized by the following function:
\begin{lstlisting}
Definition |*extract_feasible_basis*| x (Hextr: is_extreme x (polyhedron A b)) := [...]: feasible_basis A b.?\vskip.5ex?
Lemma |*extract_feasible_basis_point_of_basis*| x (Hextr: is_extreme x (polyhedron A b)) :
  x = point_of_basis (extract_feasible_basis Hextr).
\end{lstlisting}
Now, let $x^* \in \R^n$ and $y^* \in \R^p$ be such that $z^* = (x^*,y^*)$. As a consequence of the previous discussion, we can build the feasible basis \C$bas'$ of~\ref{eq:lp} from a proof that $x^*$ is an extreme point of $\Pcal(A,b)$. This proof is obtained by exploiting the fact that $\scalar{c_\Ext}{z^*} = M_\Ext$ and that $z^*$ is extreme in $\Pcal(A_\Ext,b_\Ext)$, see \C$Lemma |*extremality_ext*|$ in \C$simplex.v$. The extremality of $z^*$ naturally follows from the application of \C$Lemma feasible_point_of_basis_is_extreme$  to the feasible basis \C$bas$ of \ref{eq:phaseI}.

The previous discussion precisely describes the way we have implemented the function \C$|*pointed_simplex*|$, which completely solves the linear program~\ref{eq:lp} under the pointedness assumption.

\subsection{The General Case} In general, we can always reduce to the pointed case by showing that \ref{eq:lp} is equivalent to the following linear program in which the original variable $x \in \R^n$ is substituted by $v-w$ with $v,w \geq 0$:
\begin{equation}\tag{$\LP_\text{Pointed}$}\label{eq:lpEquiv}
\begin{array}{r@{\quad}l}
\text{minimize} & \scalar{c}{(v-w)}  \\[\jot]
\text{subject to} & A (v - w) \geq b \, , \; v \geq 0 \, , \; w \geq 0 \, , \;(v,w) \in \R^{n+n}
\end{array}
\end{equation}
The feasible set of~\ref{eq:lpEquiv} is pointed because of the constraints $v,w \geq 0$. Thus, we can apply to it the function \C$pointed_simplex$ of the previous section. In this way, we define the function \C$|*simplex*|$, which is able to solve any linear program~\ref{eq:lp}. It is implemented in \C$Section |*General_simplex*|$ of \C$simplex.v$. Its correctness proof is formalized by means of the following inductive type:
\begin{lstlisting}
Inductive |*simplex_spec*| : simplex_final_result -> Type :=
| !*Infeasible*! d of (dual_feasible_dir A d /\ '[b, d] > 0):
		simplex_spec (!*Simplex_infeasible*! d)
| !*Unbounded*! p of [/\ (p.1 \in polyhedron A b), (feasible_dir A p.2) & ('[c,p.2] < 0)]:
		simplex_spec (!*Simplex_unbounded*! p)
| !*Optimal_point*! p of [/\ (p.1 \in polyhedron A b), (p.2 \in dual_polyhedron A c) & '[c,p.1] = '[b,p.2]]: 
		simplex_spec (!*Simplex_optimal_point*! p).?\vskip.5ex?
Lemma |*simplexP*| : simplex_spec simplex.
\end{lstlisting}

In other words, when \C$simplex$ returns a result of the form \C$!*Simplex_infeasible*! d$, then \C$d$ is a certificate of infeasibility of \ref{eq:lp}, see \C$Lemma infeasibleP$ in Sect.~\ref{subsec:FeasAndFarkas}. 

Similarly, the unboundedness of the linear program \ref{eq:lp} is characterized by the fact that \C$simplex$ returns a result of the form \C$!*Simplex_unbounded*! (x,d)$. In this case, the point \C$x$ belongs to the polyhedron $\Pcal(A,b)$, and \C$d$ is a feasible direction along which the objective function strictly decreases (\C$'[c,d] < 0$). Equivalently, we can define a predicate corresponding to this situation, and prove that it is correct, as follows:
\begin{lstlisting}
Definition |*unbounded*| := if simplex is !*Simplex_unbounded*! _ then true else false.?\vskip.5ex?
Lemma |*unboundedP*| : reflect (forall M, exists y, y \in polyhedron A b /\ '[c,y] < M) unbounded.
\end{lstlisting}
Given any \C$M$, the certificate \C$y$ is built by taking a point of the form $\C$x$ + \lambda \C$d$$, where $\lambda \geq 0$ is sufficiently large.

Finally, when \C$simplex$ returns \C$!*Simplex_optimal_point*! (x,u)$, this means that \C$x$ is an optimal solution of \ref{eq:lp}, and \C$u$ is a dual feasible element which certificates its optimality (by means of \C$Lemma weak_duality$ of Sect.~\ref{subsec:reduced_costs}, since \C$'[c,x] = '[b,u]$). Thanks to this, we can define in an effective way the fact that \ref{eq:lp} admits an optimal solution, in which case we say that it is \emph{bounded}, and define its optimal value as follows:
\begin{lstlisting}
Definition |*bounded*| := if simplex is !*Simplex_optimal_point*! _ then true else false.
\end{lstlisting}
\begin{lstlisting}
Definition |*opt_value*| := if simplex is !*Simplex_optimal_point*! (x,_) then '[c,x] 
                        else 0. ((*(* not used *)*)) 
\end{lstlisting}
The correctness of these functions is given by:
\begin{lstlisting}
Lemma |*boundedP*| :
  reflect ((exists x, x \in polyhedron A b /\ '[c,x] = opt_value) /\ (forall y, y \in polyhedron A b -> opt_value <= '[c,y])) bounded.
\end{lstlisting}
As expected, this means that \C$bounded$ holds if and only if there is a point \C$x$ of $\Pcal(A,b)$ with value \C$opt_value$, and any other point of $\Pcal(A,b)$ has a value which is greater than or equal to \C$opt_value$.

\section{Outcome of the Effective Approach}\label{sec:outcome}

Several duality results immediately follow from the correctness statements of the simplex method and the resulting predicates \C$feasible$, \C$unbounded$ and \C$bounded$. In order to present them, we assume that the variables \C$A:'M_(m,n)$, \C$b:'cV_m$ and \C$c:'cV_n$ have been declared, and we define
\begin{lstlisting}
Definition |*dualb*| := (col_mx (col_mx c (-c)) (0:'cV_m)).
\end{lstlisting}
\ie, \C$dualb$ represents the vector appearing in the right-hand side of Eq.~\eqref{eq:dual}. We also define \C$dualA$ as  in Sect.~\ref{subsec:FeasAndFarkas}. In this way, \C$feasible dualA dualb$ is equivalent to the fact that the dual polyhedron $\Qcal(A,c)$ is feasible:
\begin{lstlisting}
Lemma |*dual_feasibleP*| :  reflect (exists u, u \in dual_polyhedron A c) (feasible dualA dualb).
\end{lstlisting}
Then, the second part of Th.~\ref{th:strong_duality} is given by:
\begin{lstlisting}
Theorem |*strong_duality_primal_dual_feasible*| :  feasible A b -> feasible dualA dualb ->
  exists x, exists u, [/\ x \in polyhedron A b, u \in dual_polyhedron A c  & '[c,x] = '[b, u]].
\end{lstlisting}
The remaining cases of Th.~\ref{th:strong_duality}, \ie, the cases in which one of the two linear programs is infeasible, are described by:
\begin{lstlisting}
Lemma |*strong_duality_primal_feasible_dual_infeasible*| :
	feasible A b -> ~~(feasible dualA dualb) -> unbounded A b c.?\vskip.5ex?
Lemma |*strong_duality_primal_infeasible_dual_feasible*| : 
	~~(feasible A b) -> feasible dualA dualb -> unbounded dualA dualb (-b).
\end{lstlisting}
(recall that the objective function of \ref{eq:dual_lp} is $x \mapsto \scalar{-b}{x}$). 
We also obtain the following well-known form of Farkas' Lemma characterizing the logical implication between linear inequalities:
\begin{lstlisting}
Lemma |*farkas_lemma_on_inequalities*| z : feasible A b ->
	(forall x, x \in polyhedron A b -> '[c,x] >= z) <->
		(exists u, [/\ u >=m 0, A^T *m u = c & '[b,u] >= z]).
\end{lstlisting}
All these results, which are obtained in a few lines of proof, can be found in the file \C$duality.v$ of \CoqPolyhedra.

The membership to the convex hull of a finite set of points is another property which can be defined in an effective way in our framework. Recall that a point $x \in \R^n$ belongs to the convex hull of a (finite) set $\{v^i\}_{1 \leq i \leq p} \subset \R^n$ if there exists $\lambda \in \R^p$ such that $x = \sum_{i = 1}^p \lambda_i v^i$, $\lambda \geq 0$ and $\sum_i \lambda_i = 1$. The latter constraints define a polyhedron over $\lambda \in \R^p$, and the membership of $x$ amounts to the fact that this polyhedron is feasible. This is how we arrive at the following definition of the Boolean predicate \C$|*is_in_convex_hull*|$ in the file \C$minkowski.v$ of \CoqPolyhedra, where the set $\{v^i\}_{1 \leq i \leq p}$ is formalized as a matrix \C$V:'M_(n,p)$ with columns $v^i$:
\begin{lstlisting}
Let |*e*| := (const_mx 1) : 'cV_p. ((*(* vector with constant entry 1 *)*)) ?\vskip.5ex? 
Definition |*is_in_convex_hull*| (x : 'cV_n) := 
	let Ax := col_mx (col_mx (col_mx V (-V)) (col_mx e^T (-e^T))) 1%:M in
	let bx := col_mx (col_mx (col_mx x (-x)) (col_mx 1 (-1))) (0:'cV_p) in
		feasible Ax bx.?\vskip.5ex?
Lemma |*is_in_convex_hullP*| (x : 'cV_n) :
  reflect (exists lamb:'cV_p, [/\ (lamb >=m 0), '[e, lamb] = 1 & x = V *m lamb]) (is_in_convex_hull x).
\end{lstlisting}
The \emph{separation result} states that if $x$ does not belong to the convex hull of $\{v^i\}_{1 \leq i \leq p}$, then there is a hyperplane \emph{separating} $x$ from $\{v^i\}_{1 \leq i \leq p}$. This means that $x$ is located on one side of the hyperplane, while the points $v^i$ are on the other side. 
We establish this result as follows:
\begin{lstlisting}
Theorem |*separation*| (x : 'cV_n) : ~~(is_in_convex_hull x) ->
	exists c, [forall i, '[c, col i V] > '[c, x]].
\end{lstlisting}
The certificate \C$c$ is built as \C$(dsubmx (usubmx (usubmx d)))-(usubmx (usubmx (usubmx d)))$ 
where \C$d$ is the infeasibility certificate of the polyhedron over $\lambda \in \R^p$. Our proof of the separation result reduces to technical manipulations of block matrices, as the one allowing us to define \C$c$ in terms of the infeasibility certificate.

Finally, Minkowski's Theorem states that any bounded polyhedron is equal to the convex hull of its vertices. We recover this result as the extensional equality of the predicates \C$polyhedron A b$ and \C$is_in_convex_hull matrix_of_vertices$, where \C$|*matrix_of_vertices*|$ is the matrix whose columns are the basic points of $\Pcal(A,b)$:
\begin{lstlisting}
Theorem |*minkowski*| : bounded_polyhedron A b ->
	polyhedron A b =i is_in_convex_hull matrix_of_vertices.
\end{lstlisting}
Here, \C$|*bounded_polyhedron*| A b$ is the Boolean predicate given by the disjunction of the negation of \C$feasible A b$ and the conjunction of \C$(bounded A b ei) && (bounded A b -ei)$ for all \C$i:'I_n$, where \C$ei:=(delta_mx i 0):'cV_n$ is the \C$i$th vector of the canonical base of $\R^n$. Equivalently, this means that $\Pcal(A,b)$ is bounded in the $\ell_1$-norm.
The most difficult part of \C$Theorem minkowski$ is proven in a few lines: if $x \in \Pcal(A,b)$ does not belong to the convex hull of the basic points, \C$Theorem separation$ exhibits a certificate $c$ such that $\scalar{c}{x} < \scalar{c}{x^I}$ for all feasible bases $I$ of $\Pcal(A,b)$. However, the program \C$pointed_simplex$ is able to provide an optimal feasible basis for \ref{eq:lp}, \ie, a basis $I^*$ satisfying $\scalar{c}{x^{I^*}} \leq \scalar{c}{x}$. This yields a contradiction.

\section{Conclusion}

We have presented a formalization of convex polyhedra in~\Coq{}. Its main feature is that it is based on an implementation of the simplex method, leading to an effective formalization of the basic predicates over polyhedra. We have illustrated the outcome of this approach with several results of the theory of convex polyhedra. 

Our implementation of the simplex method, especially that of Phase~I, closely follows the presentation of this method made by Schrijver in~\cite[Section~11.1]{Schrijver86}. The main difference is that we have chosen to implement the lexicographic rule~\cite{Dantzig1955} instead of Bland's rule (this choice is motivated by future work, as explained below). In contrast, the way we define the basic properties of polyhedra by means of the simplex method is non-standard, and is driven by the intuitionistic nature of the logic in \Coq{}. This leads us to derive the proof of the main mathematical statements on polyhedra from the correctness proof of the simplex method, while most textbooks follow the reversed approach (for instance,  the duality theorem is proved by means of high-level arguments based on convex analysis, and the correctness proof of the simplex method is derived from this theorem). This makes it  difficult to compare pen-and-paper proofs of the results considered here with their formalization (which consists, according to \C$coqwc$, of $2\, 784$ lines of proof). 

As a future work, we plan to deal with faces, which is a central notion in the combinatorial theory of polyhedra (early steps of an effective definition of faces are already available in the file \C$face.v$ of \CoqPolyhedra{}). The simplex method should also greatly help us to prove adjacency properties on faces, in particular, properties related with the connectivity of the (vertex-edge) graph of polyhedra. Another direction of work is to exploit our library to certify computational results on polyhedra, possibly on large-scale instances. A basic problem is to formally check that a certain polyhedron (defined by inequalities) is precisely the convex hull of a certain set of points. This is again a problem in which the simplex method with the lexicographic rule plays a central role~\cite{Avis2000,Avis1992}. The high-level nature of the data structures used in \MathCompShort{} (unary integers, abstract fields, several layers of dependent types, \etc{}) forbids any computational experiment. However, we have already done promising experiments with low-level data structures (\eg{}, based on efficient implementations of large numbers such as \C$BigN$, \C$BigZ$, \C$BigQ$) which indicate that \Coq{} may scale to large instances as we target. It remains to translate our formally proven statements to these ``computation oriented'' data structures. In this respect, we plan to investigate the approach based on refinements, like the one of~\cite{Cohen2013}. 

\section*{Acknowledgements}

The authors are very grateful to A.~Mahboubi for her help to improve the presentation of this paper, and to G.~Gonthier, F.~Hivert and P.-Y.~Strub  for fruitful discussions. The second author is also grateful to M.~Cristi\'a for introducing him to the topic of automated theorem proving. The authors thank the anonymous reviewers for their suggestions and remarks. An abridged version of this work appeared in the proceedings of ITP'17. The authors finally thank the referees of the ITP paper for their comments and suggestions.

\end{document}